\documentstyle[11pt,paspconf,epsf]{article}
\input psfig
\begin{document}

\title{The Gas Dynamics of Shell Galaxies}
\author{F. Combes and V. Charmandaris}
\affil{Observatoire de Paris, DEMIRM, Paris, F-75014, France}



\section{Introduction}

It is widely accepted that shell galaxies may form as a result of a
merger between an elliptical and a small disk galaxy. Simulations of
the {\em stellar component} have shown that the shells or ripples are
created either by ``phase-wrapping'' of debris on nearly radial orbits
(Quinn 1984, \apj, 279, 596), or by ``spatial-wrapping'' of matter in
thin disks (Dupraz \& Combes 1987, A\&A, 185, L1; Hernquist \& Quinn
1989, \apj, 342, 1). Stars, often modeled as collisionless particles,
which were originally bound to the merging companions are liberated
and oscillate with different periods around the new common central
potential. Before they relax and due to their different periods of
oscillation, they accumulate near the apocenters of their orbits, to
form shell-like features.

The fact that this scenario is indeed occuring has been questioned by
the recent observations of neutral hydrogen in systems containing
shells.  Schiminovich et al (1994, \apjl, 432, L101; 1995, \apjl, 444,
L77) have detected diffuse HI gas shells displaced just to the outside
of the stellar ripples of Centaurus~A and NGC\,2865.  These
observational results were a priori surprising, since we believe that
shells are phase-wrapped stellar structures and that the diffuse
gaseous and stellar components do not have the same behavior when
approaching the center of the potential well, in quasi-radial orbits
(Weil \& Hernquist 1993, \apj, 405, 142).  However, as it was pointed
out (Kojima \& Noguchi 1997, \apj, 481, 132) the details of the gas
dynamics depend strongly on the specific model adopted for the
interstellar medium, and modeling of the gas using smooth particle
hydrodynamics may not be very appropriate.

\section{Numerical Simulations and Discussion} 

Our numerical simulations (Combes \& Charmandaris, 1998, A\&A, in
preparation) indicate that these new HI observations can actually be
accommodated within the standard picture for the formation of shell
galaxies (see Fig.~1). We model the sinking disk galaxy using a
realistic distribution of its stellar and gaseous component, taking
into account dynamical friction and a proper treatment of the
dissipation of the gas (using a cloud-cloud collision code). During
the process of merging the tidal forces affect the gaseous component
first since it is less bound than the stars, which in turn are
liberated later. Due to the effect of dynamical friction the stars
loose energy as the galaxy spirals inwards in the potential of the
elliptical galaxy. By modifying the dissipation of the gas we can
easily reproduce the observed spatial distribution of gaseous and
stellar shells.

\begin{figure}
\centerline{\psfig{file=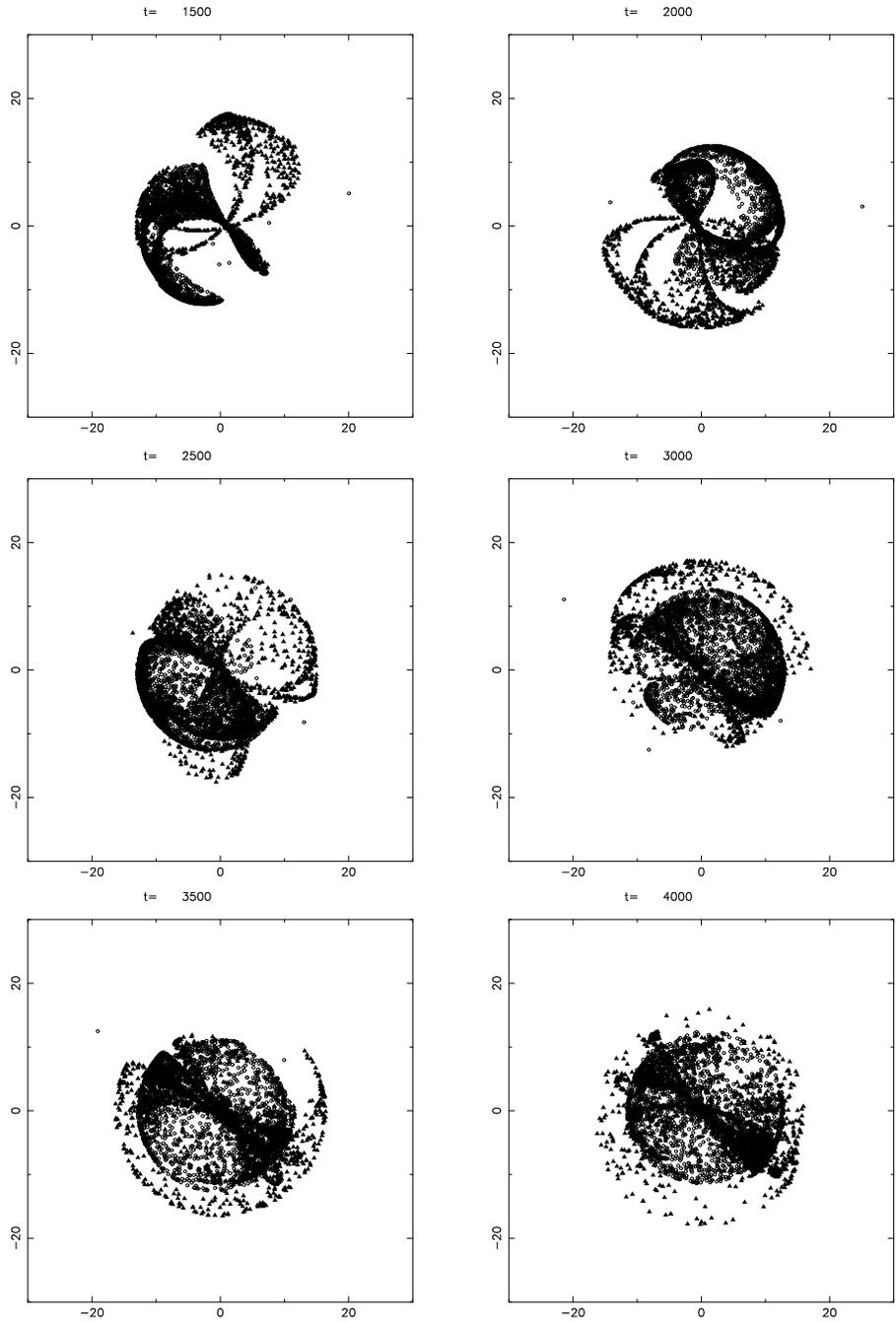,angle=180,width=12cm}}
\caption{A typical simulation result of the tidal disruption of the disk
galaxy. Stars are marked with open circles and gas clouds with filled
triangles. Note the displacement between gas and stars.}
\end{figure}

\end{document}